\documentclass{jpsj-suppl}

\title{Neutrino-Deuteron Reactions in the $\Delta$(1232) Region}

\author{T.-S. H. \textsc{Lee}$^{1}$ }

\inst{$^{1}$Physics Division, Argonne National Laboratory,
Argonne, Illinois 60439, USA}
\email{lee@phy.anl.gov}

\recdate{December 30, 2015}

\abst{ Abstract
The results from an investigation of the incoherent single pion electroweak production on the deuteron in 
the $\Delta$(1232) region are reported. The importance of including
the nucleon-nucleon final state interactions
 in extracting the neutrino-nucleon cross sections from the data obtained from the
experiments on
 the deuteron target is demonstrated.  }

\kword{neutrino-deuteron reactions}

\begin{document}
\maketitle

\section{Introduction}

It has been well recognized that the neutrino($\nu$)-nucleus reactions in the about 1-3 GeV region 
are dominated by
the excitation of the $\Delta$(1232) resonance.
To determine the neutrino properties from the data of $\nu$-nucleus 
reactions, it is necessary to develop  theoretical models for calculating the nuclear effects
in this resonance region.
Our approach in this direction consists of  two steps: (1) develop a model which can describe the
cross sections of electroweak single pion production on proton ($p$) and neutron ($n$),
(2) use the $\nu$-nucleon  reaction amplitudes generated from the constructed model  to calculate the nuclear
effects by using the $\Delta$-hole model which had been well-developed in
the studies of pion-nucleus and photon-nucleus reactions.
The objective of this report is to review the progress we have made\cite{sl1,sl2,sl3,hl,wsl}
in the step 1. The results from the step 2 had been published in Ref.\cite{nu-nucl1},
but will not be covered in this contribution. 

Since there exists no neutron target, the available data on 
$\nu$-neutron reactions were extracted from analyzing
the data  
obtained from the experiments on the  deuteron target, as done in analyzing the data from
Argonne National Laboratory (ANL), Brookhaven National Laboratory (BNL) and
European Organization for Nuclear Research (BEBC-CERN)
\cite{cam73,bar79,rad82,kit86,kit90,all80,all90,jones89}.
The essential assumption of these analyses is that
in the region near the peak of the quasi-free nucleon knock out process,
one of the nucleons in the deuteron does not participate in the reaction mechanism and can
be treated as  a spectator in evaluating the cross sections on the deuteron target.
For analyzing future experiments, it is essential to  examine the extent to which this spectator approximation
procedure is valid. In this contribution, we report on our results\cite{wsl} from investigating
this problem.

In section 2, we briefly describe our model for the nucleon.
The results on the deuteron target are presented in section 3.
A summary is given in section 4.

\section{Dynamical model for the electroweak pion production on the nucleon}

A model for investigating the pion production from $\nu$-nucleus reactions must  
first describe successfully all of the following processes on the proton $(p)$ and
neutron $(n)$: 
\begin{eqnarray}
\pi+ p(n)&\rightarrow& \pi+  N\,, \label{eq:pi-n}\\
\gamma+ p(n) &\rightarrow& \pi+  N\,, \label{eq:g-n}\\
e+p(n) &\rightarrow& \pi + N\,,\label{eq:e-n}\\
\nu +p(n) &\rightarrow& l + \pi +N\,,\label{eq:nu-n} \\
\bar{\nu}+ p (n) &\rightarrow& l + \pi +N\,,\label{eq:bnu-n}
\end{eqnarray}
where $l$ denotes the outgoing leptons, and the final $\pi + N$ can be of 
any $\pi N$ state
permitted by the electroweak conservation laws within the Standard Model.
In Refs.\cite{sl1,sl2,sl3}, we have developed such a  model
by constructing  a 
Hamiltonian of the following form
\begin{eqnarray}
H&=& H_0 + H_1 + H_{em} + H_{cc}
\label{eq:h}
\end{eqnarray}
with
\begin{eqnarray}
H_1 &=&[v_{\pi N, \pi N} + \Gamma_{\pi N,\Delta}]+[h.c.]\,,\label{eq:h1} \\
H_{em}&=& 
[v_{,\pi N,\gamma N} + \Gamma_{ \gamma N,\Delta}] +[h.c.]\,,
\label{eq:eq:hem}\\
H_{cc} &=&
[v_{\pi N, W^{\pm} N} + \Gamma_{W^{\pm} N,\Delta}]+[h.c.]\,,
\label{eq:eq:hcc}
\end{eqnarray}
where $H_0$ is the free Hamiltonian for all of the particles
 in Eqs.(\ref{eq:pi-n})-(\ref{eq:bnu-n}), $H_1$, $H_{em}$ and $H_{cc}$  describes the
$\pi N$ scattering Eq.(\ref{eq:pi-n}), 
the electromagnetic
processes Eqs.(\ref{eq:g-n})-(\ref{eq:e-n}) induced by the  photon ($\gamma$),
 and the weak charged current ($cc$) processes Eqs.(\ref{eq:nu-n})-(\ref{eq:bnu-n}) induced by the $W^{\pm}$ bosons, respectively.
In the above equations, $[h.c.]$ denote the hermitian conjugates of the interaction 
terms within the square brackets.
The two-body interactions $v_{\pi N,\pi N}$, $v_{\pi N, \gamma N}$, and
$v_{\pi N,W^{\pm} N}$ are determined
by the tree-diagrams of the well known phenomenological Lagrangian with $N$, $\Delta$,
$\pi$, $\rho$, and $\omega$ fields and the associated 
  electroweak currents. The vertex interactions $\Gamma_{\pi N,\Delta}$,
$ \Gamma_{ \gamma N,\Delta}$, and $\Gamma_{W^{\pm} N,\Delta}$ are determined by fitting
the data of the processes Eq.(\ref{eq:pi-n}), Eqs.(\ref{eq:g-n})-(\ref{eq:e-n}), and
Eqs.(\ref{eq:nu-n})-(\ref{eq:bnu-n}), respectively.

With the Hamiltonian defined above, the reaction amplitudes on the nucleon with
a $\pi N$ final state
 can be written\cite{sl1} as
\begin{eqnarray}
T_{\pi N,\alpha N} (E)&=& t_{\pi N,\alpha N}(E) +\bar{\Gamma}_{\pi N}(E)
\frac{1}{E-m^0_\Delta-\Sigma(E)}\bar{\Gamma}^\dagger_{\alpha N}(E)\,,
\label{eq:t-n}
\end{eqnarray}
 where $m^0_\Delta = 1299$ MeV is the bare mass of the $\Delta$, and 
$\alpha N= \pi N, \gamma N, W^{\pm}N$.
The non-resonant $\pi N \rightarrow \pi N$ amplitude $t_{\pi N,\pi N}(E) $ and the dressed
$\Delta \rightarrow \pi N$ vertex $\bar{\Gamma}_{\pi N}(E)$ are  defined by 
\begin{eqnarray}
t_{\pi N,\pi N}(E) &=& v_{\pi N,\pi N}[1+\frac{1}{E-H_0+i\epsilon}t_{\pi N,\pi N}(E)]\,,\\
\bar{\Gamma}_{\pi N}(E) &=&[1+ t_{\pi N,\pi N}(E)\frac{1}{E-H_0+i\epsilon}]\Gamma_{\pi N,\Delta} \,,\\
\bar{\Gamma}^\dagger_{\pi N}(E) &=&\Gamma^\dagger_{\pi N,\Delta}
[1+ \frac{1}{E-H_0+i\epsilon}t_{\pi N,\pi N}(E)]\,.
\end{eqnarray}
Keeping only the first order in electroweak coupling, the amplitudes for the electroweak processes
are accordingly defined by
\begin{eqnarray}
t_{\pi N,\beta N}(E) &=&[1+t_{\pi N,\pi N}(E)\frac{1}{E-H_0+i\epsilon}]v_{\pi N,\beta N}\,,\\
\bar{\Gamma}^\dagger_{\beta N}(E)&=& \Gamma^\dagger_{\beta N,\Delta}+
\bar{\Gamma}^\dagger_{\pi N}(E) \frac{1}{E-H_0+i\epsilon}v_{\pi N, \beta N}\,,
\end{eqnarray}
where $\beta N =\gamma N, W^{\pm} N$.
The $\Delta$ self-energy in Eq.(\ref{eq:t-n}) is defined by
\begin{eqnarray}
\Sigma(E)= \bar{\Gamma}^\dagger_{\pi N}(E)\frac{1}{E-H_0+i\epsilon}\Gamma_{\pi N,\Delta}.
\end{eqnarray}

\begin{figure}[htbp] \vspace{-0.cm}
\begin{center}
\includegraphics[width=0.5\columnwidth]{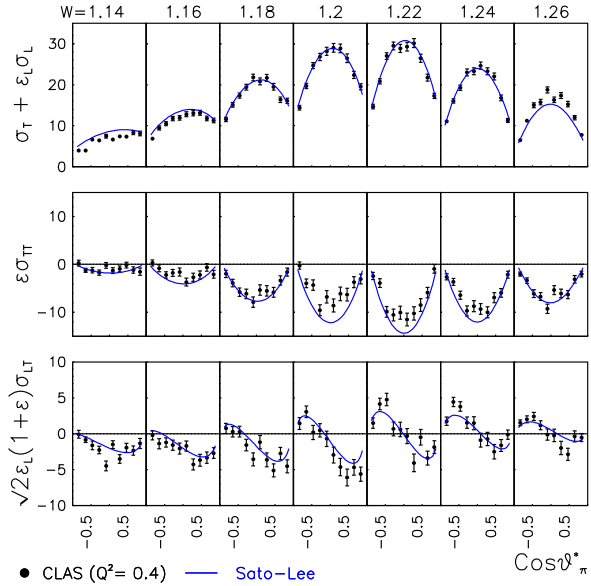}
\caption{Predictions of Refs.\cite{sl1,sl2} are compared with
 the structure function data of $p(e,e'\pi^0)p$. The data are from Ref.\cite{joo}}
 \label{fg:eepi}
\end{center}
\end{figure}

\begin{figure}[htbp] \vspace{-0.cm}
\begin{center}
\includegraphics[width=0.4\columnwidth]{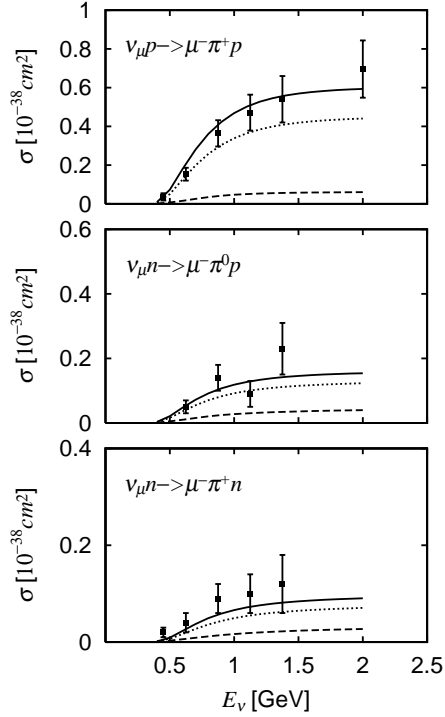}
\caption{The total cross sections of $\nu$-nucleon reactions.
The solid curves are from the full calculations of Ref.\cite{sl3}.
The dotted curves are from the calculations including only the contributions from the $\Delta$ resonant
amplitudes.
The dashed curves are from the calculations including only the contributions from the non-resonant amplitudes.
The data are from Ref.\cite{nu-tot}}
 \label{fg:nutot}
\end{center}
\end{figure}

 The parameters of the model have been determined by fitting all of the available data of the
processes Eqs.(\ref{eq:pi-n})-(\ref{eq:bnu-n}) in  the $\Delta(1232)$ energy region.
The overall good fits to the data 
have been given in Refs.\cite{sl1,sl2,sl3} as well as in many experimental papers on
the electromagnetic processes Eqs.(\ref{eq:g-n})-(\ref{eq:e-n}).
Here we only give two examples.
In Fig.\ref{fg:eepi} we show that the predicted
 structure functions of $p(e,e'\pi^0)p$ are in good agreement with the data.
Another example is shown in Fig.\ref{fg:nutot}
for the
fits (solid curves) to the available total cross section data of $\nu+ N \rightarrow l+\pi +N$. 
We note here that the calculated total cross sections for the neutron target
are not in good agreement with the data. 
 Before we improve our model, it is however necessary to examine the extent to which the
spectator approximation
 used in extracting these
data  from the data on the deuteron target is valid, as discussed in section 1.
We address this important question  in the next section.


\section{Electroweak reactions on the deuteron}

\begin{figure}[htbp] \vspace{-0.cm}
\begin{center}
\includegraphics[width=0.3\columnwidth]{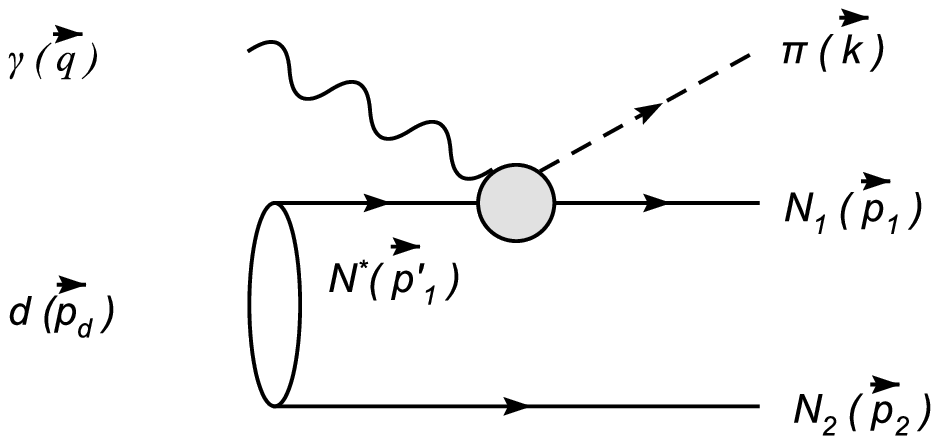}
\includegraphics[width=0.3\columnwidth]{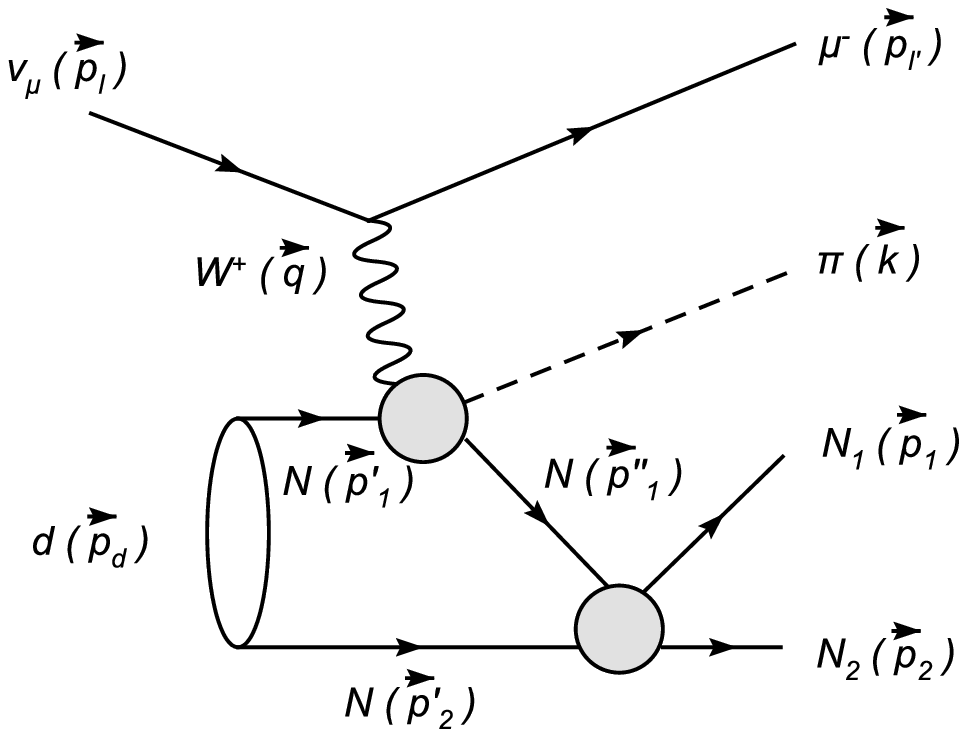}
\includegraphics[width=0.3\columnwidth]{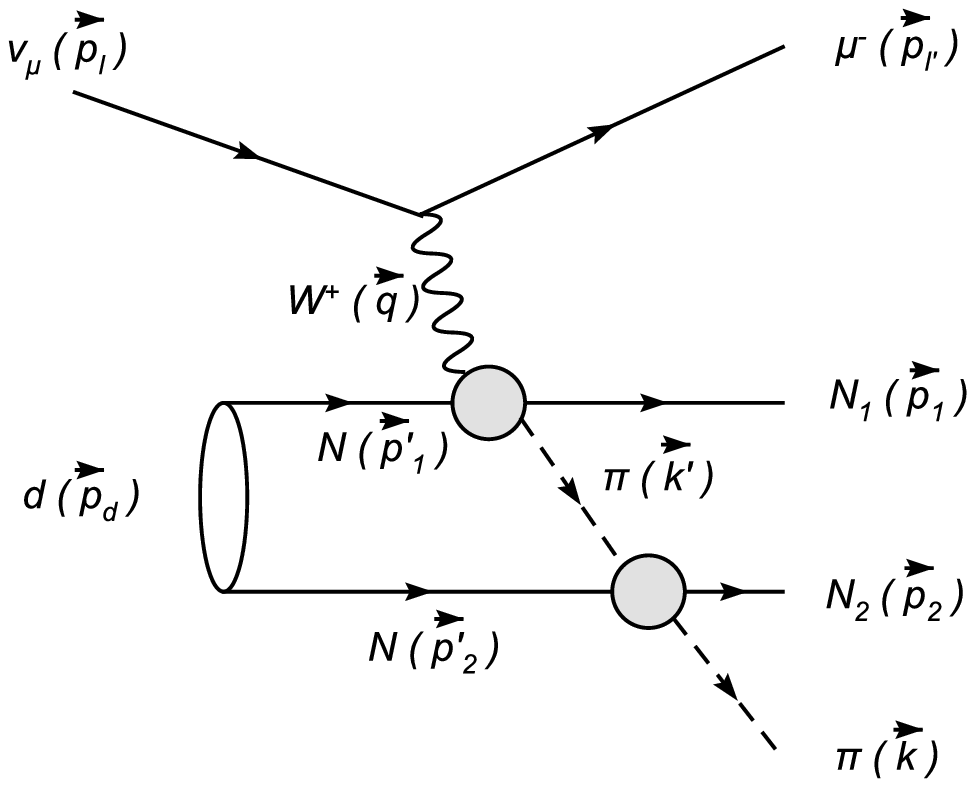}
\caption{The impulse (left), $NN$ final state interaction (center), and $\pi N$ final
state interaction (left)  mechanisms of 
$\nu (p_{l}) + d (p_d) \rightarrow l' (p_{l'}) + \pi (k) + N_1 (p_1) + N(p_2)$}.
 \label{fg:mech}
\end{center}
\end{figure}

In our investigations\cite{hl,wsl}, the amplitudes of the
incoherent electroweak pion production on the deuteron consist of an impulse ($imp$) term, 
a nucleon-nucleon ($NN$) final state interaction term, and
a pion-nucleon ($\pi N$) final state interaction term: 
\begin{eqnarray}
T_{\pi NN,\beta d}=<\pi [N_1N_2]_A | J^{(Imp)}(E)+ J^{(NN)}(E)+J^{(\pi N)}(E)|\beta \Phi_d>\,,
\label{eq:t-d}
\end{eqnarray}
where $\beta= \gamma, W^{\pm}$, $\Phi_d$ is the deuteron bound state wavefunction, and
$<\pi [N_1N_2]_A|$ is a plane-wave $\pi NN$ state with an anti-symmetrized $NN$  component.
 Each term in Eq.(\ref{eq:t-d}) can be
calculated from the single-nucleon matrix elements $T_{\pi N,\beta N}$  defined in section 2,
and the $NN$ scattering amplitude $T_{NN, NN}(E)$ which can be generated from various
realistic $NN$ potentials.
Schematically, they can be written as 
\begin{eqnarray}
J^{(Imp)}(E) &=& [T_{\pi N_1, \beta N_1}(E_{\pi N_1})]+[1\rightarrow 2] \label{eq:imp}\,, \\
J^{(NN)}(E)&=& [T_{N_1,N_2, N_1,N_2}(E_{N_1N_2})\frac{|\pi N_1 N_2><\pi N_1 N_2|}
{E- H_0+i\epsilon} T_{\pi N_1, \beta N_1}(E_{\beta N_1})]+ [1\leftrightarrow 2] \label{eq:nn},\\
J^{(\pi N)}(E)&=& [T_{\pi,N_2, \pi,N_2}(E_{\pi N_2})\frac{|\pi N_1 N_2><\pi N_1 N_2|}
{E- H_0+i\epsilon} T_{\pi N_1, \beta N_1}(E_{\beta N_1})]+ [1\leftrightarrow 2]\,, \label{eq:pin}
\end{eqnarray}
where $E_{ab}$ is  the energy associated with particles $a$ and $b$, as specified
in Ref.\cite{wsl}.
In Fig.\ref{fg:mech}, we illustrate each term in  the calculations of the amplitudes of the
$\nu (p_{l}) + d (p_d) \rightarrow l' (p_{l'}) + \pi (k) + N_1 (p_1) + N(p_2)$ process.

\subsection{Test of the model in the study of $\gamma + d \to N+N+\pi $ reactions}
\begin{figure}[htbp] \vspace{-0.cm}
\begin{center}
\includegraphics[width=0.45\columnwidth]{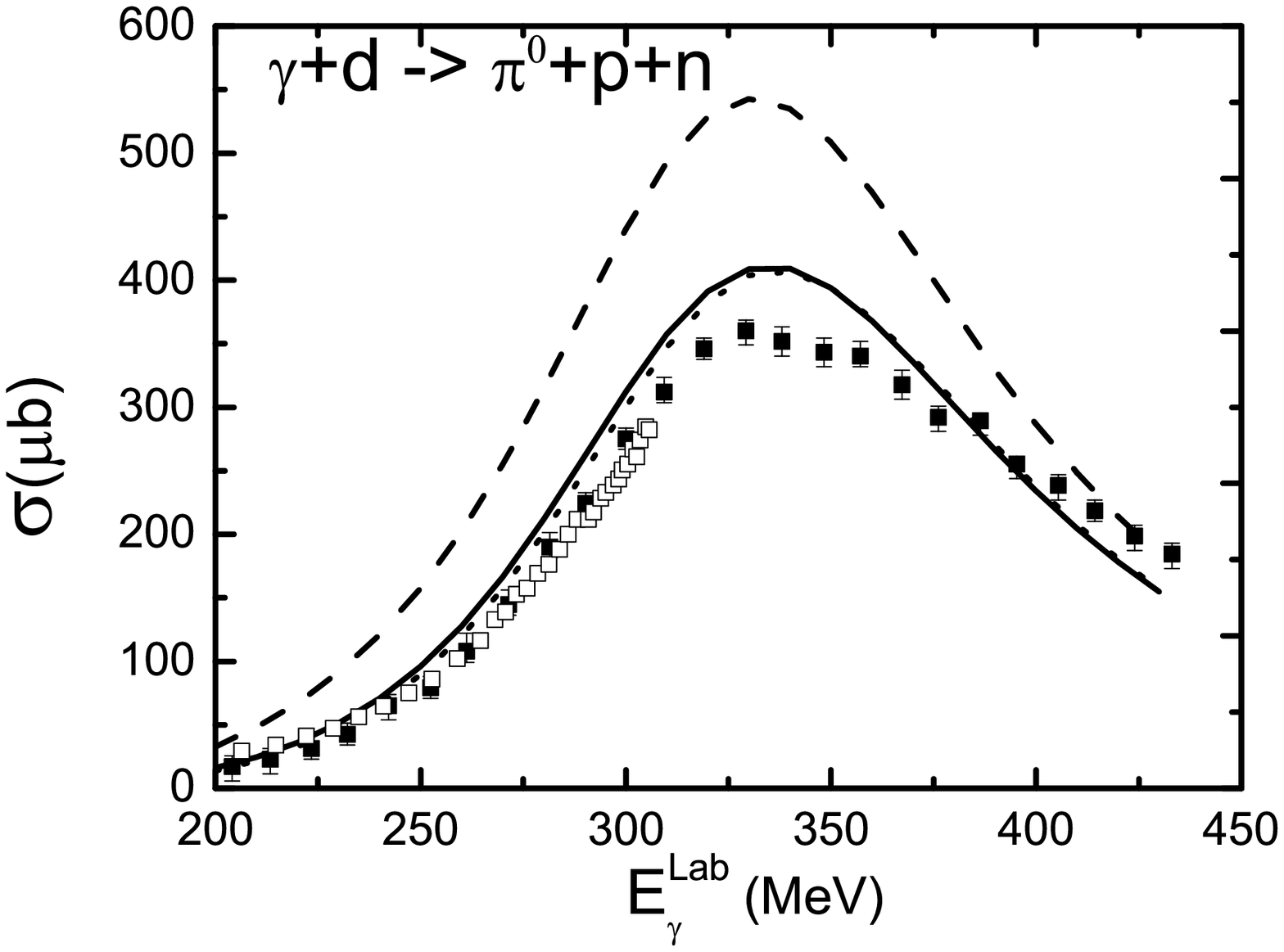}
\includegraphics[width=0.4\columnwidth]{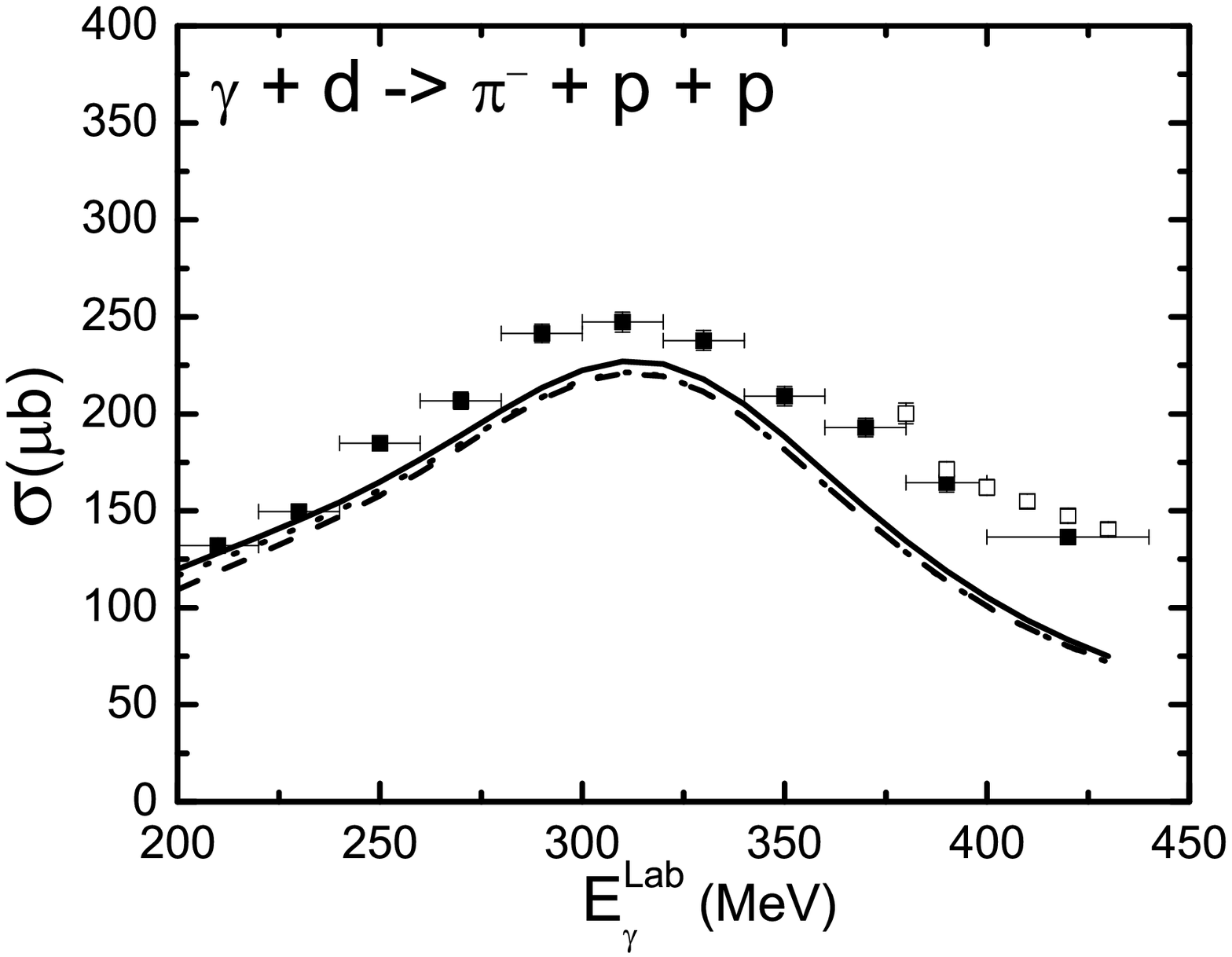}
\caption{The total cross sections of $\gamma + d \to \pi^0 + n + p$.
The  dashed,  dotted, and solid
curves  are from the  calculations include
only the impulse term , the impulse + ($NN$ final state interaction),
and the Impulse + ($NN$ final state interaction) + ($\pi N$ final state interaction), respectively.
Data are from Refs. ~\cite{krus,benz,asai}.}
 \label{fg:pi0tot}
\end{center}
\end{figure}

We first test our calculation procedures by comparing the calculated cross sections
of $\gamma + d \to N+N+\pi $ reactions with the data.
Our results for the total cross sections
of $\gamma+ d \rightarrow \pi^0+n+ p$ are shown in the left-side of  Fig.\ref{fg:pi0tot}.
 When only the
impulse term $J^{(Imp)}$ is included,
we obtain the  dashed curve. It is greatly reduced
to  dot-dashed curve when the $np$ final state interaction term $J^{(NN)}$
is added in the calculation. When the $\pi N$ final state interaction
term $J^{(\pi N)}$ is also included in our full calculation, we obtain the solid curve.
Clearly, the $np$ re-scattering effects are very large while the $\pi N$
re-scattering give negligible contributions.

Similar comparisons for the total cross sections
of the $\gamma+ d \rightarrow \pi^- +p+p$ process
are shown in the right side of Fig.\ref{fg:pi0tot}.
 Here we see that both the $pp$ and $\pi N$
final state interactions are weak.
 Comparing the two results  shown in Fig.\ref{fg:pi0tot},
we see the
large differences between $np$ and $pp$ final state interactions.

This results shown in Fig.\ref{fg:pi0tot} are consistent with the results of
Refs.\cite{darwish,fix,lev06,sch10,tarasov}, and 
 can be understood qualitatively from the properties of
 the initial deuteron wave function
and the final $NN$ wave functions.
We first observe that the final $\pi NN$ interactions are mainly due to  the s-wave
$NN$ states in the considered energy region.
For $\pi^0np$ final state, the dominant
final $np$ state is $^3S_1+^3D_1$ which has the same quantum number as
the initial deuteron state.
 Since the radial wave functions of the deuteron and the scattering state in this
partial wave must
be orthogonal to each other,
one expects that the loop integrations over these two wave functions
 are strongly suppressed
compared with those from the impulse approximation calculations.
 On the other hand, there is no such orthogonality relation for the
 $^1S_0$ $pp$ in the $\pi^-pp$.
Consequently the final state interaction effect in
the $\gamma+ d \rightarrow \pi^0+n+p$ is much stronger than that
in the $\gamma +d \rightarrow\pi^-+ p+p$.

We see in
Fig.\ref{fg:pi0tot}
that  our full calculations (solid curves) are in reasonable
agreement with the data in both the shapes and magnitudes, while some improvements are still needed
in the future.  Thus
our approach based on Eqs.(\ref{eq:t-d})-(\ref{eq:pin}) is valid for 
predicting the $\nu+ d \rightarrow \mu +\pi+ N+ N$
cross sections,  as given in the next subsection.

\subsection{Predictions of $\nu + d \rightarrow \pi + N + N$ cross sections } 

In section 1, we describe
 a spectator approximation which was used in the
previous analyses\cite{cam73,bar79,rad82,kit86,kit90,all80,all90}
to extract the neutrino-induced single pion production
cross sections on the proton and neutron from the data on the deuteron target.
Here we  use our model to examine the extent to which this
approximation is valid.

\begin{figure}[htbp] \vspace{-0.cm}
\begin{center}
\includegraphics[width=0.49\columnwidth]{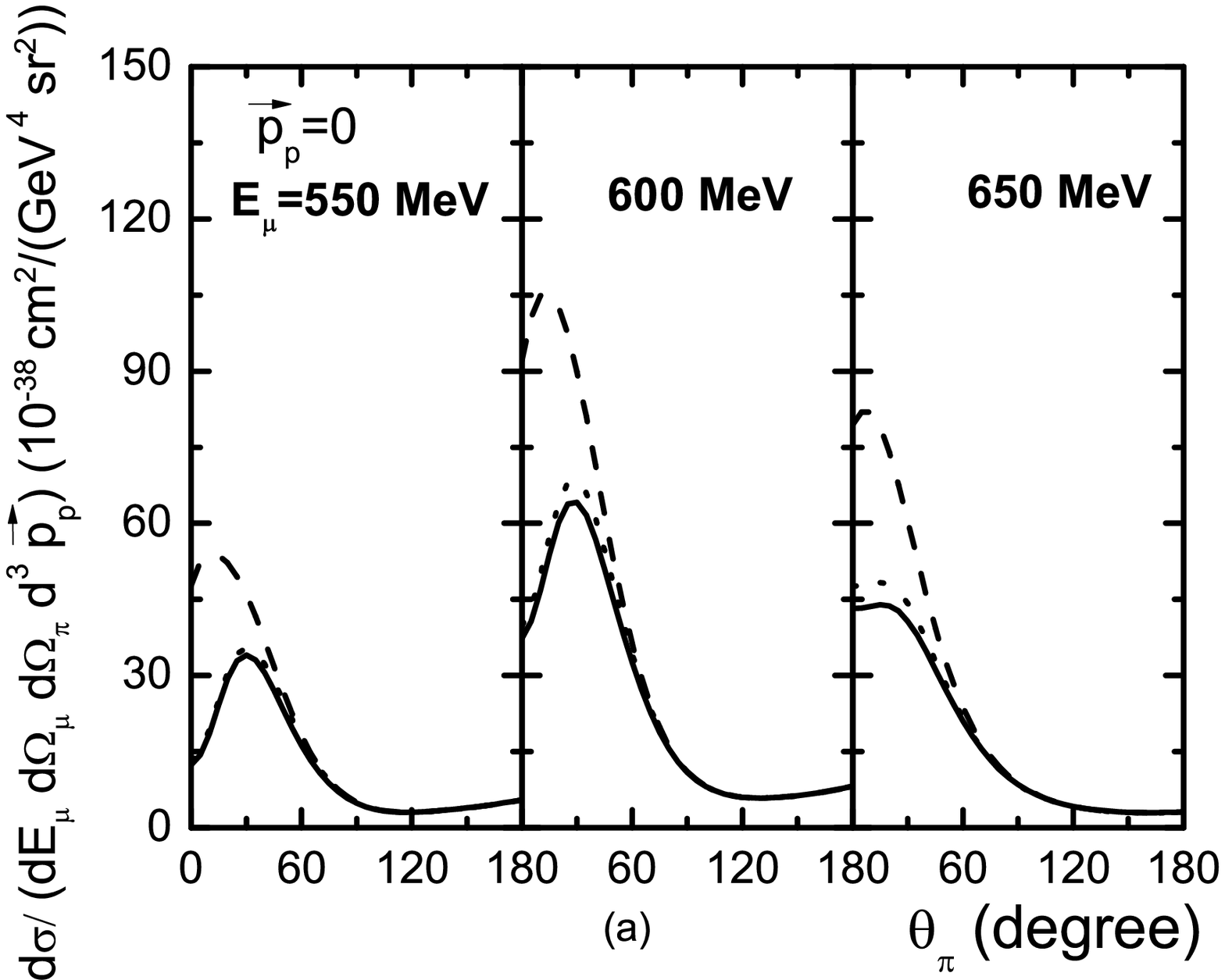}
\includegraphics[width=0.49\columnwidth]{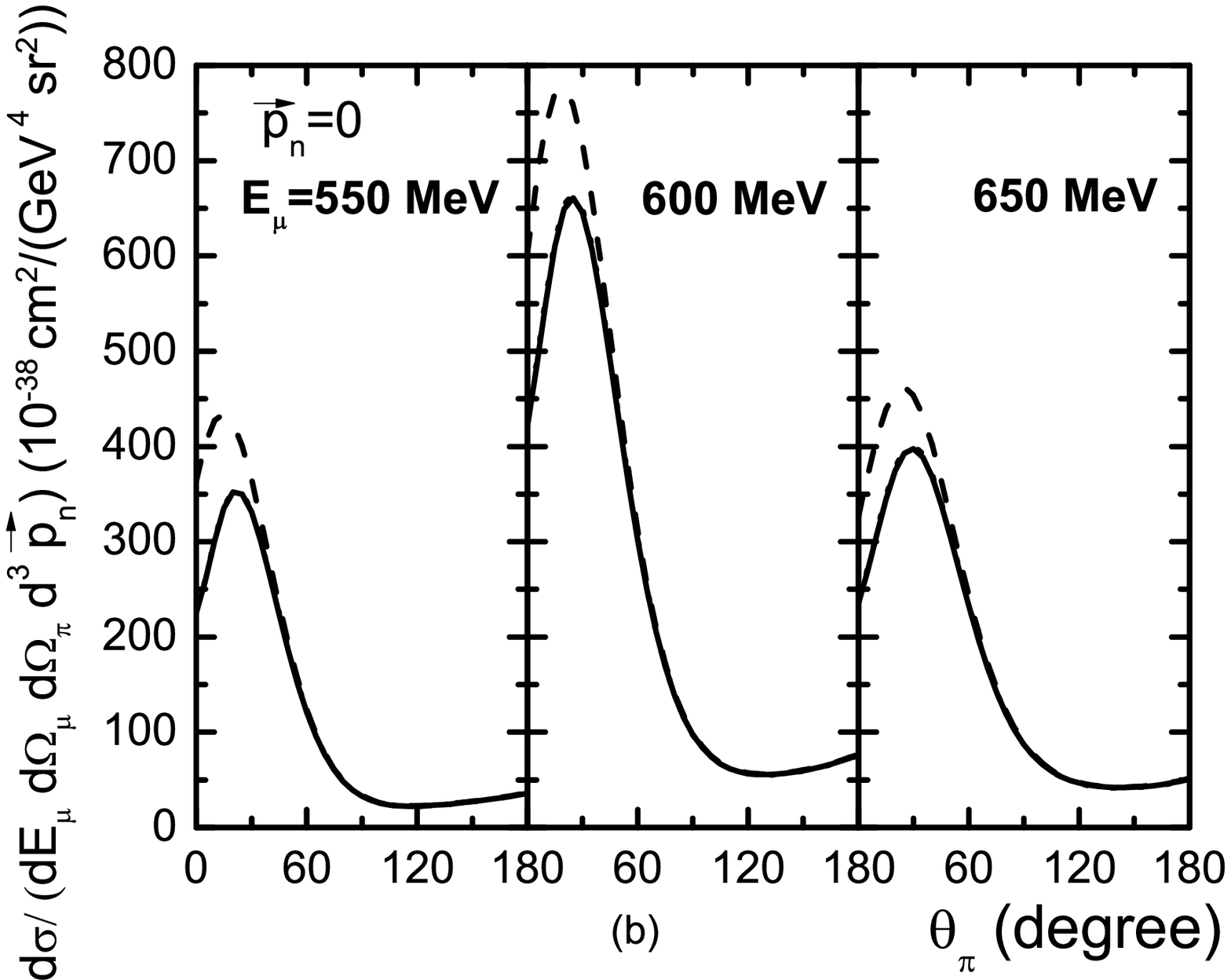}
\caption{ The differential cross sections
$d\sigma / dE_{\mu^-} d\Omega_{\nu_\mu\,\mu^-} d\Omega_{\pi^+} d^3\vec{p}_N$
of $\nu_\mu + d \to \mu^- + \pi^+ + p + n$
as function of pion scattering angle in the laboratory system.
The left (right) figure is for the
proton $\vec{p}_p=0$ (neutron $\vec{p}_n=0$) spectator kinematics.
The outgoing muon energy is $E_{\mu^-}$=550, 600, and 650MeV.
The  dashed, dotted, and  solid  curves are from
calculations including only the Impulse term, Impulse +($NN$ final state interaction),
and Impulse +($NN$ final state interaction)+($\pi N$ final state interaction), respectively.
The  dotted and  solid curves are almost indistinguishable since the
$\pi N$ final state interaction effects are very small.
}
 \label{fg:pnzero}
\end{center}
\end{figure}

To be specific, we consider the case that
the spectator nucleon is at rest. If there is no final state interactions,
the  $\nu + d \rightarrow l^- +\pi^+ + n+p$ cross section
is only from  the pion production on the other nucleon  which is also at rest in the
deuteron rest frame. Then the
 cross sections measured at  the kinematics that the final proton (neutron)
 at rest $\vec{p}_p=0$ ($\vec{p}_n=0$)  are simply the cross sections of
$\nu_\mu + n \to \mu^- + \pi^+ + n$ ($\nu_\mu + p \to \mu^- + \pi^+ + p$ ).
These
 are the dashed curves in Fig. \ref{fg:pnzero}.
When the $NN$ final-state interaction terms are included, we obtain the dotted curves in
Fig. \ref{fg:pnzero}. The solid curves are obtained  when
the $\pi N$ final state interaction is also
included in the calculations.
Clearly, the $NN$ re-scattering can significantly change the cross sections while the
$\pi N$ re-scattering  effects are weak.
It is also important to note that
the $NN$  re-scattering effects on the cross sections for $\vec{p}_p=0$
are rather different that for $\vec{p}_n=0$.

The results shown in Fig. \ref{fg:pnzero} strongly suggest that the
spectator assumption used in
the previous analyses\cite{cam73,bar79,rad82,kit86,kit90,all80,all90} is
not valid for the $\pi^+$ process $\nu + d \rightarrow \mu^-+ \pi^++ n+p$.
This result is due to
the large $np$ re-scattering effects, as explained in the subsection 3.1.

We have also examined the results for $p_s=0$ for the
$\pi^0$ process $\nu + d \rightarrow \mu^-+ \pi^0+ p+p$.
Here we find that the spectator assumption is a good approximation for
extracting the cross section on the nucleons from the deuteron target.
This is of course due to the weak $pp$ final state interactions,
as can be seen in the right side of Fig.\ref{fg:pi0tot}.

\section{Summary}

We have studied the incoherent  electroweak pion production reactions on the deuteron target. 
It is found that the predicted $\gamma+ d \rightarrow \pi^0+ n+p,\,\,\, \pi^-+p+p$ cross sections
agree well with the available data. The cross sections of
 $\nu+  d \rightarrow l^-+ \pi^++ n+p, \,\,\, l^-+ \pi^0+ p+p $ have been predicted.
It is demonstrated that the $NN$ final state interactions have large effects on
the processes with  $np$ final states. Our results indicate the need of
re-analyzing the available data on the deuteron target to extract more accurately
the cross sections on the
neutron and proton targets.

\vspace{1cm}
This work was supported by the U.S. Department of Energy, Office of Nuclear Physics Division,
under Contract No. DE-AC02-06CH11357. 
This research used resources of the National Energy Research Scientific Computing Center,
which is supported by the Office of Science of the U.S. Department of Energy
under Contract No. DE-AC02-05CH11231, and resources provided on Blues and/or Fusion,
high-performance computing cluster operated by the Laboratory Computing Resource Center
at Argonne National Laboratory.

\end{document}